\documentclass[pra,twocolumn,aps,showpacs,nofootinbib,floatfix,groupedaddress]{revtex4-1}
\usepackage{amsmath}
\usepackage{amssymb}
\usepackage{graphicx}
\usepackage{verbatim}
\usepackage[colorlinks=true,linkcolor=blue,citecolor=blue,urlcolor=blue]{hyperref}
\usepackage{txfonts}
\usepackage{xr}

\begin{document}

\title{{Terahertz field control of interlayer transport modes in cuprate superconductors}}
\author{Frank Schlawin$^1$}
\email{frank.schlawin@physics.ox.ac.uk}
\author{Anastasia S. D. Dietrich$^1$}
\author{Martin Kiffner$^{1,2}$}
\author{Andrea Cavalleri$^{1,3}$}
\author{Dieter Jaksch$^{1,2}$}
\affiliation{$^{1}$ Clarendon Laboratory, University of Oxford, Parks Road, Oxford OX1 3PU, United Kingdom}
\affiliation{$^{2}$ Centre for Quantum Technologies, National University of Singapore, 117543 Singapore}
\affiliation{$^{3}$ Max Planck Institute for the Structure and Dynamics of Matter, Luruper Chaussee 149, 22761 Hamburg, Germany}

\begin{abstract}

We theoretically show that terahertz pulses with controlled amplitude and frequency can be used to switch between stable transport modes in layered superconductors, modelled as stacks of Josephson junctions. We find pulse shapes that deterministically switch the transport mode between superconducting, resistive and solitonic states. We develop a 
{simple} model that explains the switching mechanism as a destablization of the centre of mass excitation of the Josephson phase, made possible by the highly non-linear nature of the light-matter coupling.

\end{abstract}

\maketitle

\begin{figure*}
\centering
\includegraphics[width=0.8\textwidth]{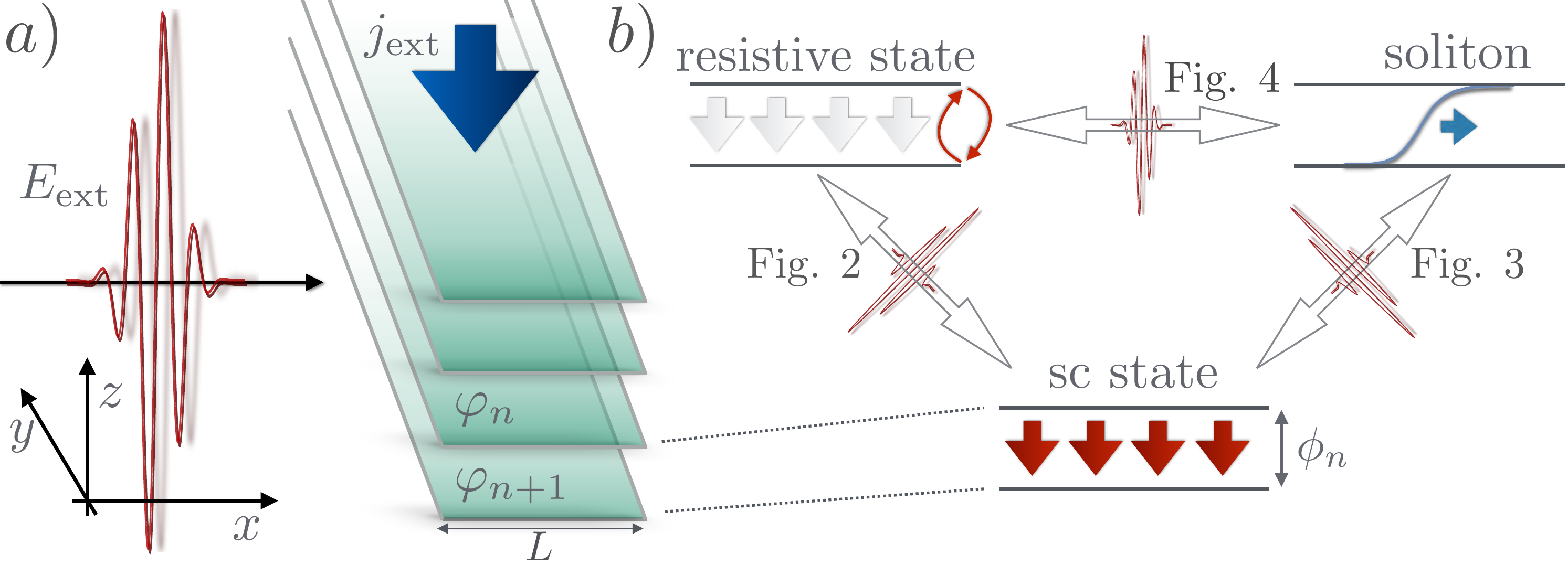}
\caption{
a) A short, layered superconductor is driven by a short laser pulse and a dc current $j_{\text{ext}}$. b) The interplay between the current, the nonlinearity of the crystal, and the light field $E_{\text{ext}}$ allows for the switching between the superconducting state, where the current flows as a supercurrent (red), the resistive state, where the current is supported by quasiparticles (grey), causing oscillatory supercurrents, and solitonic states, where quasiparticle and supercurrents coexist. Numerical results of simulations of these transitions are shown in Figs.~\ref{fig.time-evolution1}, \ref{fig.time-evolution2} and \ref{fig.time-evolution3}, respectively.
}
\label{fig.setup}
\end{figure*}

\section{Introduction}

Coherent control of the quantum dynamics in atomic or molecular systems forms an important pillar of modern quantum physics.
Recent experimental progress in the generation and detection of terahertz radiation \cite{Hirori11} expands this field from atomic ensembles to solid state devices \cite{Kampfrath11}, and opens up unprecedented possibilities for the control and manipulation of macroscopic systems through light-matter interactions \cite{Kampfrath13, Nicoletti16}.
For instance, the nonlinear driving of phonon modes allows for the manipulation of electronic degrees of freedom in solids~\cite{Forst11, Subedi14, Knap15, Sentef16, Millis17}. 
This coupling gives rise to a plethora of exciting effects like the melting of charge density waves~\cite{Forst14, Forst14b, Mankowsky17}, { the excitation of synthetic magnetic fields \cite{Nova17},} the possibility to drive metal-insulator transitions~\cite{Rini07, Liu12}, {control hetero-interfaces \cite{Caviglia12, Foerst15, Foerst17},} or even the controlled creation of transient superconductivity~\cite{Fausti11, Hu14, Mankowsky14, Mitrano16}.

A different type of nonlinearity arises in the $c$-axis electrodynamics of 
layered superconductors~\cite{Savelev05, Savelev05b, Savelev06, Bulaevski08, Lin10, Savelev10, Guarcello15,Dienst11, Dienst13, Rajasekaran15}, {which are well described by stacked, coupled Josephson junctions for temperatures sufficiently far below the critical temperature~\cite{Kleiner92}.}
These systems exhibit a nonlinear coupling between external currents and Josephson plasmons, which is routinely employed for the creation of coherent THz radiation~\cite{Ozyuzer07, Wang10, Bulaevski07, Lin08, Lin09, Tachiki10, Koshelev08, Koyama09, Nonomura09, Hu, Welp13}. The inverse process - the light control of electric currents in layered superconductors - could offer exciting prospects for future quantum technologies. For example, optical driving could assist the flow of supercurrents in presence of strong magnetic fields above $H_{c1}$.
 Materials with high critical temperature $T_c$ like cuprates could then be used for applications where strong, superconducting currents need to be sustained to create high magnetic fields as, e.g., in magnetic resonance imaging~\cite{Ansorge16}. However, this possibility of enhancing material properties by external driving has remained largely unexplored to date. 

In this paper we consider a layered superconductor (sc) consisting of stacked two-dimensional sc layers as shown in Fig.~\ref{fig.setup}a). The material is driven by light polarised along the $z$-axis {(which is parallel to the crystallographic $c$-axis)}, and the whole stack carries a dc current along the $z$-axis{, which is smaller than the critical Josephson current $j_J$}. We consider a parameter regime in which the system can occupy one of three states: The current can be transmitted either as a supercurrent, in which Cooper pairs tunnel between adjacent layers through the Josephson effect. It can also be transmitted as a quasiparticle current of individual charge carriers, with the corresponding voltage inducing plasma oscillations that can emit coherent light \cite{Hu, Welp13}. Additionally, solitonic solutions represent dynamical steady states in which quasiparticle and supercurrents coexist in the system. In contrast to \cite{Bulaevski08, Lin10}, where stimulated emission due to external radiation was discussed, this paper explores the response to light in a regime where the pulse cannot be considered a perturbation of the undriven steady state. 

We show how strong THz pulses can induce transitions between these three macroscopic quantum states. Since the plasmon dispersion depends on the macroscopic state, each state reacts differently to external driving, thus creating parameter regimes in which only one transition responds to the pulse, while other excitation paths remain ``dark".  Therefore, tailored pulses can act as deterministic switches between pairs of states. Roughly speaking, low-frequency driving destabilizes high-voltage states, while high-frequency radiation can force the system into high-voltage states.
We explain this behaviour by the light-induced destabilisation of plasma oscillations in the centre-of-mass mode. 

The paper is organized as follows: Our model and the numerical approach are introduced in section~\ref{sec.model}. In section~\ref{sec.simulations}, we present simulations of the light-induced switching between macroscopic quantum states. These are put in a broader context in section~\ref{sec.parameter}, where we explore the influence of pulse parameters, and develop a simplified toy model to explain the destabilization mechanism of an initial state. Finally, we conclude with a discussion of the relevance of our results for future experiments in section~\ref{sec.discussion}.

\section{Model}
\label{sec.model}

Our model {for the setup} shown in Fig.~\ref{fig.setup}a) follows from a description of the material in terms of the Josephson coupling between sc layers and macroscopic electromagnetism.
The polarisation of the pulses along the $z$-direction reduces the problem to dynamics along the $x$- and $z$-axis. The electromagnetic field couples to the gauge-invariant phase differences $\phi_n = \varphi_n - \varphi_{n+1} - \frac{2 \pi}{\Phi_0} \int_{n}^{n+1} \!\! dz \; A_z$ between adjacent layers. Here, $A_z$ denotes the vector potential in $z$-direction, {$\varphi_n$ the {order parameter} phase in the $n$-th layer}, and $\Phi_0$ the magnetic flux quantum. Its dynamics is coupled to the magnetic fields in the $y$-direction{, since spatial changes of $\phi_n$ along the $x$-direction translate into magnetic field perpendicular to it.}
{Throughout this manuscript, we use dimensionless units, in which case the equations of motion may be written as}
 \cite{Koshelev00, Koshelev01},
\begin{align}
\frac{\partial^2 \phi_n}{\partial \tau^2} + \nu_c \frac{\partial \phi_n}{\partial \tau} + \sin \phi_n - \frac{\partial h_n}{\partial \xi}+\eta (\xi, \tau) = j_{\text{ext}}, \label{eq.EOM1} \\
\left( \ell^2 \nabla_n^2 - 1 \right) h_n + \frac{\partial \phi_n}{\partial \xi} + \nu_{ab} \frac{\partial}{\partial \tau} \left( \frac{\partial \phi_n}{\partial \xi} - h_n \right) = 0. \label{eq.EOM2}
\end{align}
{ Here, $h_n$ denotes the dimensionless magnetic field, the damping constants $\nu_c$ and $\nu_{ab}$ are proportional to the quasiparticle conductivity along the c-axis and the xy-plane, respectively, and $\ell$ describes the strength of their magnetic coupling.
Their expression in terms of physical quantities is given in appendix~\ref{appendix.units}.}
The discrete $z$-derivative is defined as $\nabla^2_n h_n \equiv h_{n+1} + h_{n-1} - 2 h_n$. The term $\sin \phi_n$ accounts for the Josephson coupling between the layers.  
In addition, we include the random driving term $\eta (\xi, \tau)$, emulating thermal phase fluctuations { to ensure that our results are stable with respect to these fluctuations. In this paper, we focus on the low-temperature regime where these fluctuations are small. In particular, this means that they do not drive phase slips, nor do they excite thermal solitons. Any change in the macroscopic state is due to the external driving. Their }
impact on reflectance measurements is discussed in appendix~\ref{sec.fluctuations}. 
We approximate the boundary conditions for the layered structure \cite{Bulaevski07, Bulaevski06} by simple nonradiative conditions
\begin{align}
h_n (\tau) \big|_{\xi = 0} = h_{ext} (\tau),
\end{align}
and 
\begin{align}
h_n (\tau) \big|_{\xi = L} = 0, 
\end{align}
where $h_{ext}$ denotes the external pulse. {These conditions} represent excellent approximations, since the boundary electric field is suppressed by a large impedance mismatch to the vacuum \cite{Lin09}. We write the pulse as 
\begin{align}
h_{ext} (\tau) &= A e^{- ( \tau - \tau_0)^2 / (2 \sigma^2)} \sin (\omega_{dr} \tau + \gamma), \label{eq.pulse-form}
\end{align}
where $\omega_{dr}$ denotes the driving frequency, $\sigma$ the pulse duration, {$\tau_0$ the pulse delay}, { $\gamma$ its carrier envelope phase (CEP),} and $A$ its amplitude. 
Our parametrisation is chosen such that $A$ denotes the maximal phase difference $\phi$ created by the pulse at the boundary at a given time. In order to chose realistic field strengths, we estimate \cite{Rajasekaran15} for LSCCO, according to which a phase difference $\phi \sim 1$ corresponds to a pulse with peak field intensity of $20~$kV / cm, with $100~$kV / cm being within experimental reach. {Thus we remain in the parameter regime where the of validity of the mean-field model has been tested experimentally [25]. The effects of light-induced pair-breaking as well as microscopic materials details should be negligible. Furthermore, for the short pulses employed, heating effects do not play a role.}

We consider a system which is sufficiently large along the $z$-axis, such that we can neglect finite-size effects due to coupling to connecting electrodes. As pointed out in \cite{Bulaevski07}, this is the case when the number of junctions $N$ exceeds the magnetic coupling length, $i.e.$ $N \gtrsim \ell$. The external driving can then synchronise the dynamics in the junction, $i.e.$ $\phi_n \rightarrow \phi$ and $h_n \rightarrow h$. This approximation was shown to yield excellent results in the simulation of the optical response of LSCCO in \cite{Rajasekaran15}, and will also be used throughout this work. { To check this assumption further, we have run simulations with up to 20 junctions, and found that interlayer coupling does not alter the switching from superconductive to the resistive state. }

Within our model, the value of $\ell$ becomes irrelevant in this limit, and is set to $\ell = 0$ in Eq.~(\ref{eq.EOM2}). We numerically solve equations~(\ref{eq.EOM1}) and (\ref{eq.EOM2}) using the method of lines (discretising the spatial dimension), and solving the resulting coupled ordinary equations with the IDA package \cite{IDA} in Mathematica.
Furthermore, we limit our studies to the case where the interlayer voltage drop of two solitons would exceed the voltage drop of the resistive state. In this limit, only the three states shown in Fig.~\ref{fig.setup}b) exist. Increasing the length $L$ or the external current $j_{\text{ext}}$ to go beyond this limit would allow states with several solitons (but not add qualitatively new physics within our model).
 


\begin{figure*}
\centering
\includegraphics[width=0.8\textwidth]{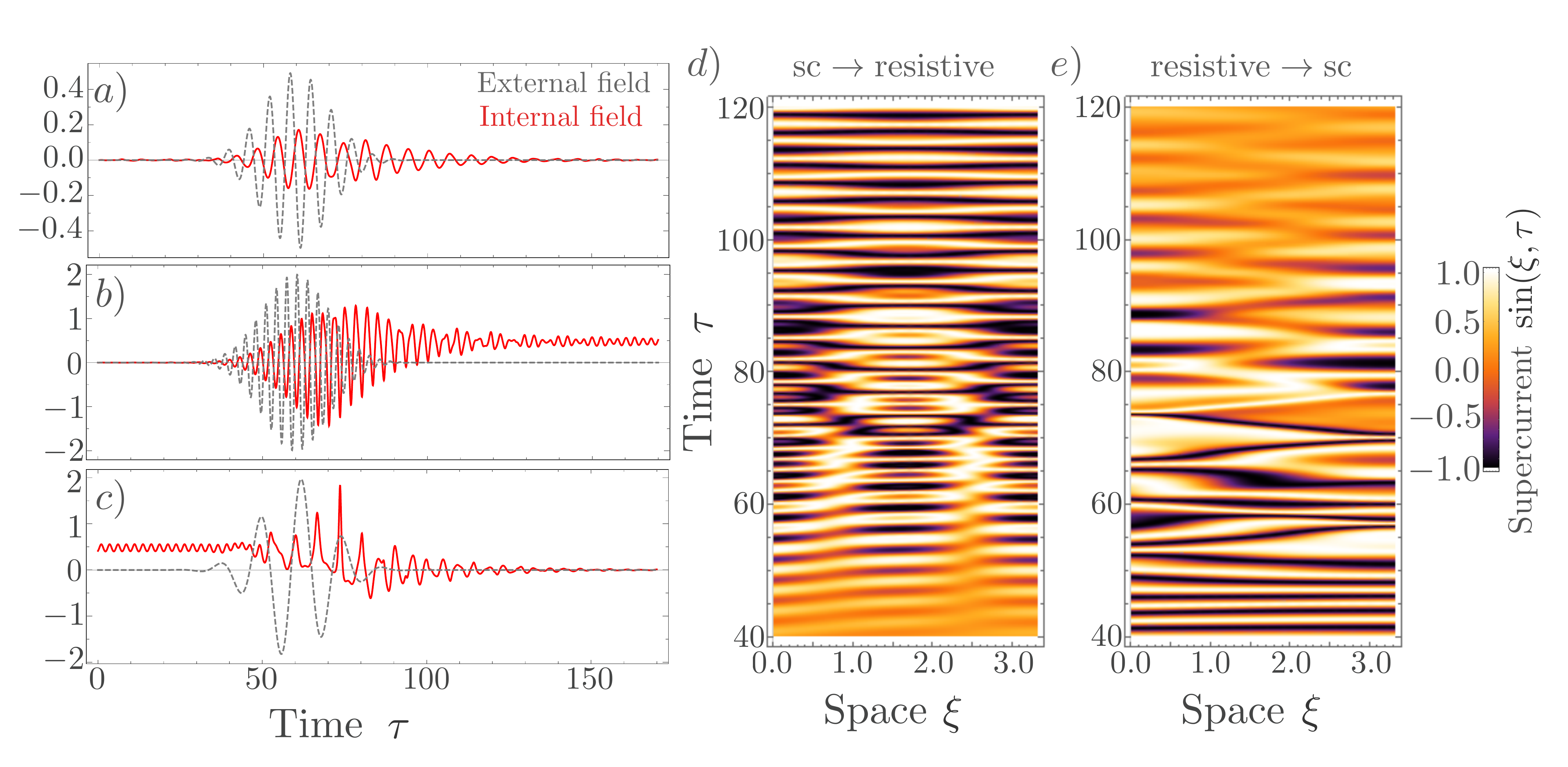}
\caption{ {Linear response and light-induced switching between sc and resistive state: }a) Electric field evolution (red) at the left boundary after excitation by a weak pulse (grey, dashed) at the plasma resonance, $i.e.$ $\omega_{dr} = 1$. The residual signal at large times stems from thermal fluctuations. 
b) Excitation {from an initial sc state} by a strong pulse with $\omega_{dr} = 2$ {and amplitude $A = 1.8$} that drives the system into the resistive state, signified by a constant electric field, $i.e.$ the emergence of a voltage drop across the junctions. c) Destabilisation of the resistive state by another strong pulse (with $\omega_{dr} = 0.5$), which disturbs the voltage drop, and thus stops the quasiparticle current. d) Supercurrent evolution $\sin (\phi (\xi, \tau) )$ in the junction during the excitation process shown in panel b). e) Supercurrent evolution $\sin (\phi (\xi, \tau))$ during the destabilisation {process shown in} panel c). 
{ We fix the values $\nu_c = \nu_{ab} = 0.1$ and $j_{\text{ext}} = 0.25$, $L = 3.3$, as well as the pulse parameters $\tau_0 = 60$ and $\gamma = 0$ in the simulations.
} 
}
\label{fig.time-evolution1}
\end{figure*}

\section{Light-induced dynamics}
\label{sec.simulations}

In this section we numerically study the light induced transitions between sc, resistive and solitonic states shown in Fig.~\ref{fig.setup}b).
We first discuss the linear response to weak pulses when the system is initialised in the sc state, given by $\phi_{\text{sc}} = \arcsin j_{\text{ext}}$  and $h = 0$. { The optical signature of the coherent Josephson coupling in this state consists of a sharp edge in the reflectivity spectrum. }Weak pulses near this plasma edge excite plasma oscillations as shown in Fig.~\ref{fig.time-evolution1}a), where we depict the external field $ h_{ext} (\tau)$ (gray) and the internal electric field {$d\phi / d\tau$}. 
The field oscillation follows the driving pulse with phase difference $\pi$, indicating the absorption of energy from the external field. The Fourier transform of the reflected field with respect to $\tau$ shows that the plasma resonance peak is located at $(1 -  j_{\text{ext}}^2)^{1/2}$ \cite{Salerno84}. {We present simulations of reflectivity signals of such weak pulses in appendix~\ref{sec.fluctuations}. In Fig.~\ref{fig.reflectivity}, the plasma edge is seen as a sharp reduction of the reflected signal frequency component. Thus, the system absorbs energy very efficiently at this frequency.}
Far from the plasma edge, $\omega_{dr} \gg 1$ or $\omega_{dr} \ll 1$, weak waves cannot penetrate the system \cite{Savelev10}. As we shall see in the following, this changes dramatically for stronger pulses, when the linear response no longer applies. The system cannot be treated as an effective medium, that is not affected by the light, and strong plasma oscillations can actively influence the state of the superconductor.

\subsection{Switching between superconducting and resistive transport}

We first focus on the transition between the sc and the resistive state. 
Fig.~\ref{fig.time-evolution1}b) shows the interaction of a strong pulse with the system in the sc state. The strong pulse excites plasma oscillations which no longer disperse, but instead build up in magnitude, and stabilise uniform plasma oscillations. This can be seen in Fig.~\ref{fig.time-evolution1}d) where we depict the supercurrent evolution {$\propto$ }$\sin \phi (\xi, \tau)$ across the entire junction, and where traveling waves can be distinguished at short times from the stable uniform oscillations after the interaction with the pulse. 
A finite voltage drop is stabilized - the constant offset of the electric field in Fig.~\ref{fig.time-evolution1}b), when averaged over the weak oscillations - satisfying the Ohmic relation $d\phi / d\tau = j_{\text{ext}} / \nu_c$.\footnote{According to the AC Josephson relation, $d\phi/ d\tau$ gives the local voltage in the junction.}
The macroscopic {state} is now given approximately by the McCumber state \cite{McLaughlin78},
\begin{align}
\phi_{\text{res}} (\tau) = \omega_0 \tau + \Im \left\{  \frac{e^{i \omega_0 \tau}}{\omega_0^2 - i \omega_0 \nu_c } \right\}, \label{eq.resistive}
\end{align}
where $\omega_0 = j_{\text{ext}} / \nu_c$. The weak oscillations on top of the offset {in Fig.~\ref{fig.time-evolution1}b)} stem from Cooper pair tunnelling induced by the ac Josephson relation, and are described by the second term in Eq.~(\ref{eq.resistive}). The pulse has thus switched the system from the sc to the resistive state.

Strong laser pulses may also be employed to destabilise the quasiparticle current by disturbing the voltage drop across the junction, and thereby switching the system 
to the sc state. An example is given in Figs.~\ref{fig.time-evolution1}c) and \ref{fig.time-evolution1}e), where a pulse 
drives the system initialised in the resistive state~(\ref{eq.resistive}). In contrast to the sc state, the resistive state does not show a plasma edge (see {appendix~\ref{appendix-EOM}}), and low-frequency waves ($\omega < 1$) can penetrate the system. 
The driving frequency is too small to directly couple to the resistive state plasma oscillations with $\omega_0 = j_{\text{ext}} / \nu_c \simeq 2.5$. However, as can be seen in Fig.~\ref{fig.time-evolution1}c), when the electric field becomes negative, it locally cancels the voltage drop, and thereby stops the quasiparticle current. This in turn destabilises the oscillations, and eventually destroys them. As one can see in panel~\ref{fig.time-evolution1}e), the first weaker oscillation of the incoming pulse at $\tau \simeq 50$ [compare with Fig.~\ref{fig.time-evolution1}b)] shifts the phase of the resistive state's uniform oscillations. The second, stronger oscillation {at $\tau \simeq 60$}  disturbs the voltage such that the fast plasma oscillations collapse, and finally decay on a time scale $\sim \nu_c^{-1}$. Thus, this pulse has switched the system back into the sc state.

\begin{figure*}
\centering
\includegraphics[width=0.8\textwidth]{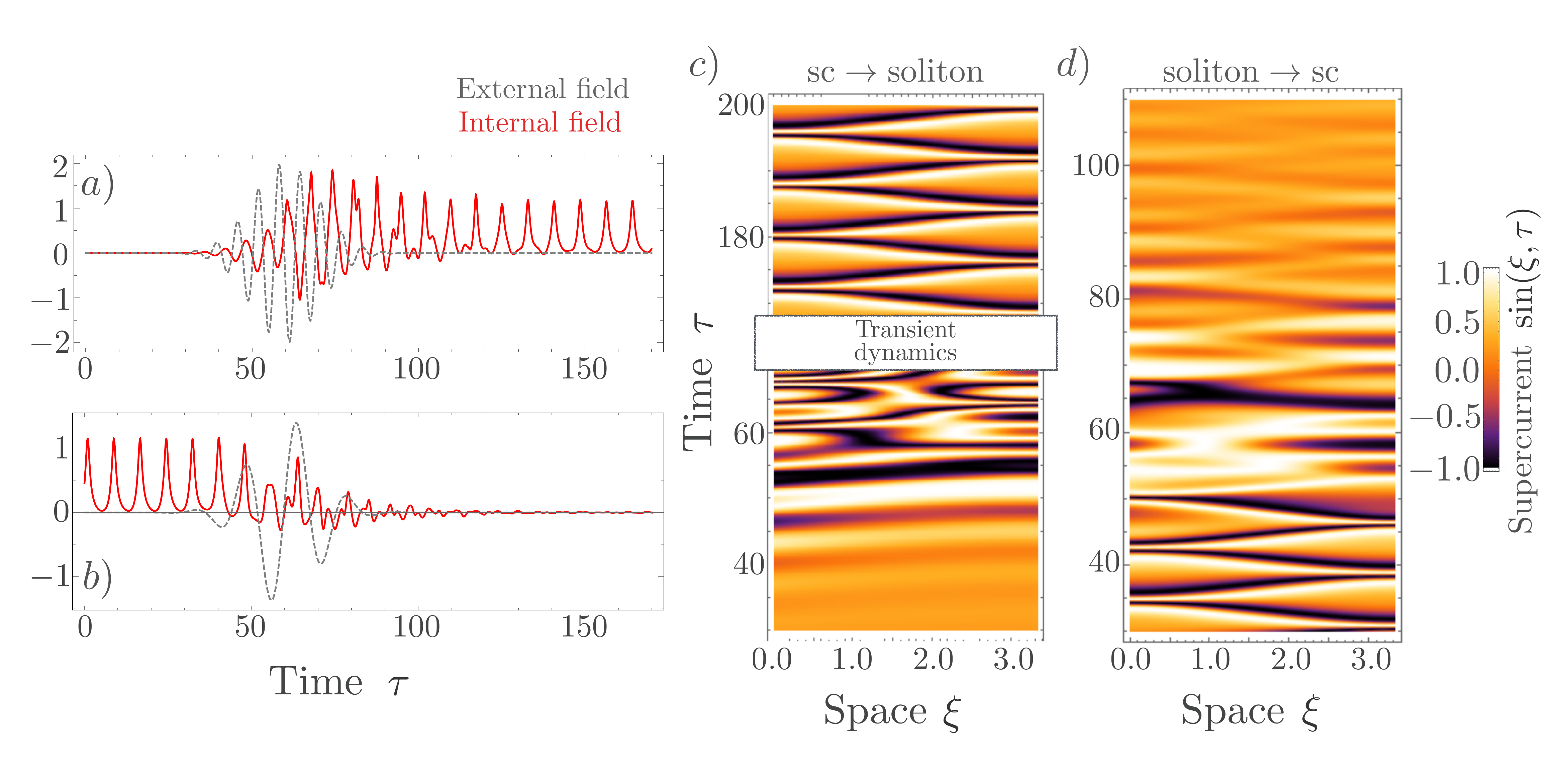}
\caption{ {Light-induced switching between sc and solitonic state: }a) Electric field evolution (red) at the left boundary {during the} excitation by a pulse with $A = 2$ and $\omega_{dr} = 1$ that drives the system from the sc into the solitonic state. b) The inverse transition from the solitonic to the sc state is induced by a pulse with amplitude $A = 1.5$ and frequency $\omega_{dr} = 0.4$.
c) Supercurrent evolution during the optical excitation of a traveling soliton by the same pulse as in panel a). d)  Supercurrent evolution during the destruction of the soliton by the same pulse as in panel b).
The remaining parameters are identical to Fig.~\ref{fig.time-evolution1}.
}
\label{fig.time-evolution2}
\end{figure*}

\subsection{Switching between superconducting and solitonic transport}

Fig.~\ref{fig.time-evolution2} 
shows the driving of the sc state by a strong pulse, which excites the solitonic state. 
Here, after a longer transient evolution which we skip in Fig.~\ref{fig.time-evolution2}c), the nonlinearity induces a traveling soliton. 
This soliton represents a different kind of stable dynamical state, in which the currents are carried by both quasiparticle and supercurrent contributions. Moving Josephson solitons {are quantum vortices of the condensate surrounded by supercurrents, which carry one magnetic flux quantum with them. }In a spatially infinite medium - in the absence of external currents and dissipation - {their wavefunction is given by} \cite{McLaughlin78} 
\begin{align}
\phi_{\text{soliton}} (\xi, \tau) &= 4 \tan^{-1} \left[ \exp \left( \pm \frac{\xi - u \tau}{\sqrt{1 - u^2}} \right) \right],\label{eq.soliton}
\end{align}
with the velocity $| u | \leq 1$. The solitonic wave, Eq.~(\ref{eq.soliton}), describes a traveling step-like increase of the phase by $2 \pi$, and thereby drives the supercurrent $\sin \phi (\xi, \tau)$ through a full cycle, see Fig.~\ref{fig.time-evolution2}a). 
It can be shown that, in a first approximation, external current and dissipation do not significantly alter this shape, but merely affect the velocity by supplying or draining kinetic energy, respectively \cite{McLaughlin78}. At equilibrium, such an analysis yields the so-called power-balance velocities,
$u_{\infty} = \pm 1 / (1 + ( 4 \nu_c /(\pi j_{\text{ext}} ) )^2 )^{1/2}$,
which in turn allow us to compute the fundamental frequency of the kink motion, $\omega_{\infty} \simeq 2 \pi \frac{u_{\infty}}{L}$.
Whenever a soliton hits the boundary, it is reflected as an anti-soliton (which also increases the phase, while moving in the opposite direction), emitting a burst of radiation \cite{Hu}, and thus providing a direct experimental fingerprint for its creation by the driving pulse. 
Averaged over one oscillation period, it amounts to a voltage drop of $V / \omega_p = \hbar \omega_{\infty} / (2 e)$, that is smaller than the voltage drop for the resistive state $\hbar \omega_0 / (2 e)$.

The soliton may also be destroyed by optical means. This is exemplified in Fig.~\ref{fig.time-evolution2}b), where a pulse disturbs the soliton, such that it disperses into plasma wavepackets which quickly decay. Similarly to the resistive state, we find the soliton to be unstable against driving below the plasma edge. 

\begin{figure*}
\centering
\includegraphics[width=0.8\textwidth]{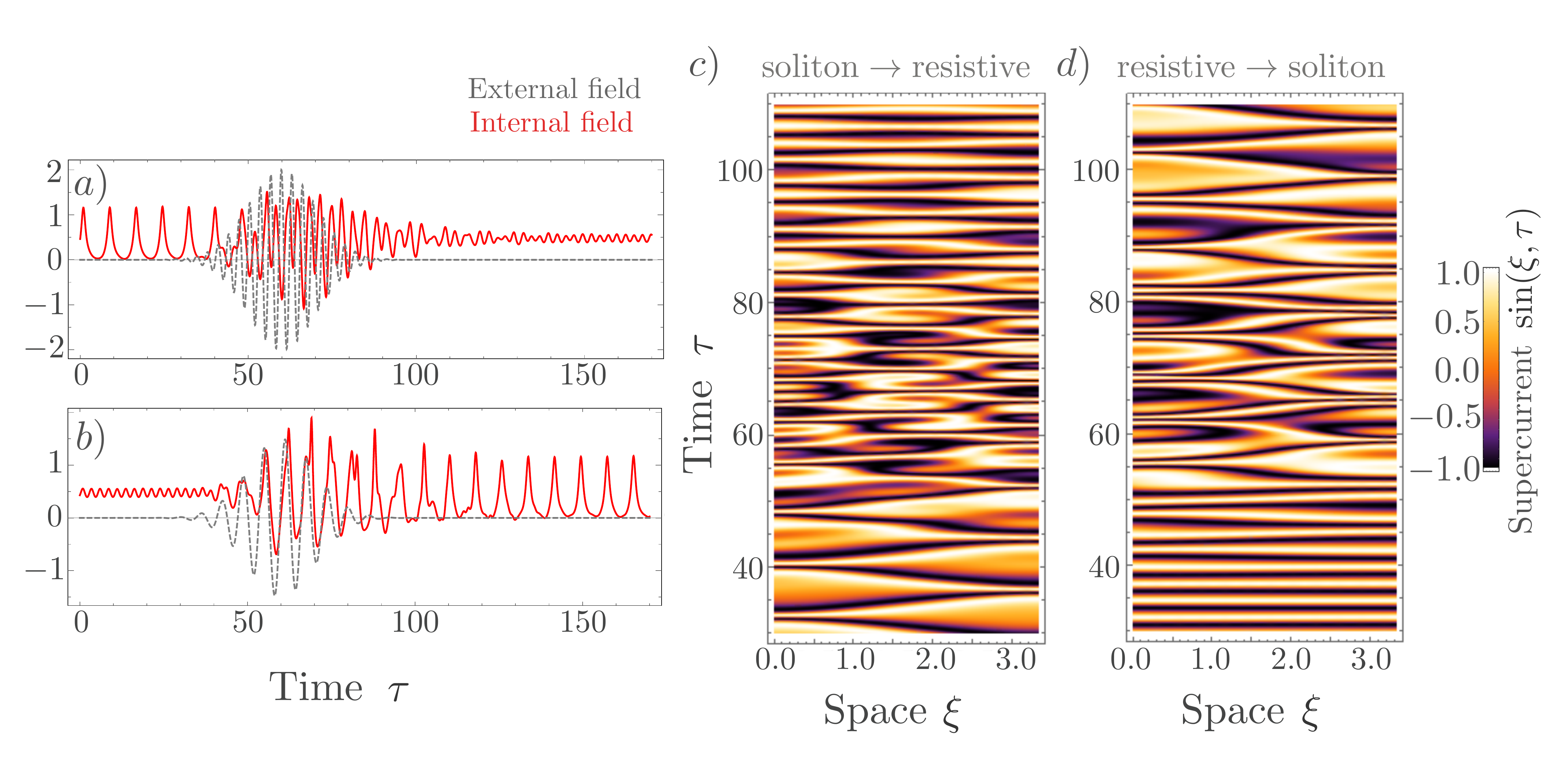}
\caption{a) Electric field evolution (red) at the left boundary after excitation by a pulse with $A = 2$ and $\omega_{dr} = 2$ ($i.e.$ with the same parameters as the pulse in Fig.~\ref{fig.time-evolution1} that switches the sc to the resistive state) that drives the system from the solitonic into the resistive state, signified by a constant electric field, $i.e.$ the emergence of a voltage drop across the junctions. b) The inverse transition from the resistive to the solitonic state is induced by a pulse with amplitude $A = 1.5$ and frequency $\omega_{dr} = 1$ { (like a pulse that creates a soliton from the sc state)}. c) Supercurrent evolution $\sin (\phi (\xi, \tau) )$ in the junction during the excitation process shown in panel a). d) Supercurrent evolution $\sin (\phi (\xi, \tau))$ during the destabilisation {shown in} panel b). {The remaining parameters are identical to Fig.~\ref{fig.time-evolution1}.}
}
\label{fig.time-evolution3}
\end{figure*}

\subsection{Switching between solitonic and resistive states}

{In Fig.~\ref{fig.time-evolution3}, we show} the switching between solitonic and resistive states. 
Figs.~\ref{fig.time-evolution3}a) and \ref{fig.time-evolution3}c) depict the interaction of a junction in the solitonic state with a strong pulse 
that switches the sc to the resistive state.
The solitonic state is signified by bursts of radiation that are emitted whenever the soliton hits the junction boundary. As can be seen in Fig.~\ref{fig.time-evolution3}a), these bursts are stopped by the pulse, and a constant voltage drop {[with small oscillations on top, see Eq.~(\ref{eq.resistive})]} is stabilised instead. In the supercurrent plot of panel~\ref{fig.time-evolution3}c), this change is reflected in the destruction of the traveling phase slip, and the emergence of uniform plasma oscillations across the entire junction. 

Conversely, a pulse with the same driving frequency that excited a soliton from the sc state can disrupt the voltage, and switch the system from the resistive into the solitonic state [see Figs.~\ref{fig.time-evolution3}b) and \ref{fig.time-evolution3}d)].

\begin{figure}
\centering
\includegraphics[width=0.49\textwidth]{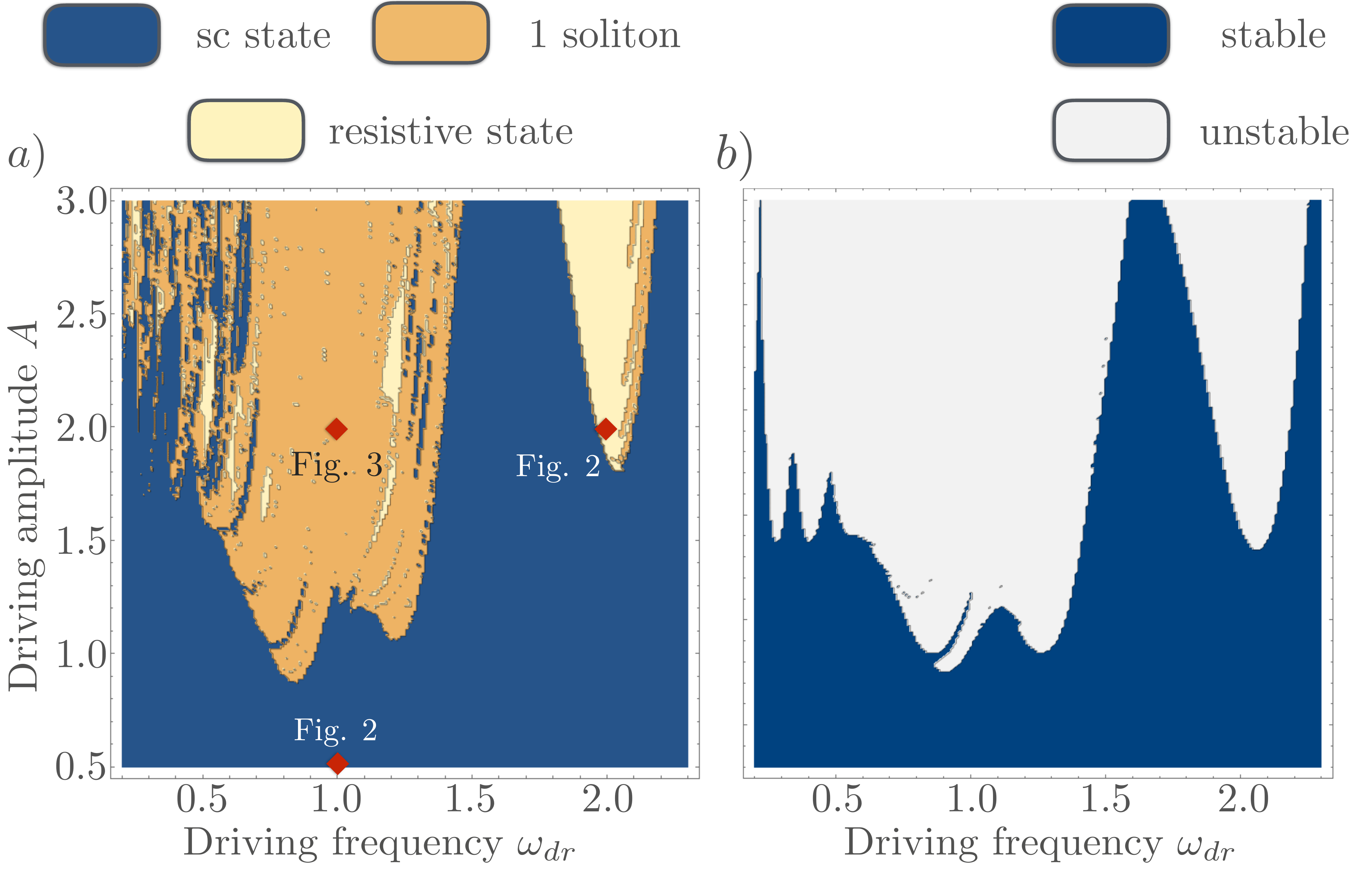}
\caption{a) Final macroscopic quantum state after excitation by a pulse with inverse bandwidth $\sigma = 10$, amplitude $A$, and carrier frequency $\omega_{dr}$. The system is initialised in the sc state. 
 The parameters of Figs.~\ref{fig.time-evolution1} and \ref{fig.time-evolution2} are indicated by red diamonds. {Blue regions indicate parameters in which the system remains in the sc state, and the pulse merely induces transient plasma waves. Dark (bright) ochre regions indicate parameters in which traveling solitons (resistive states) are excited. }
 b) Destabilisation of the COM mode according to Eq.~(\ref{eq.com-mode}).
 }
\label{fig.parameter-space}
\end{figure}

\begin{figure}
\centering
\includegraphics[width=0.49\textwidth]{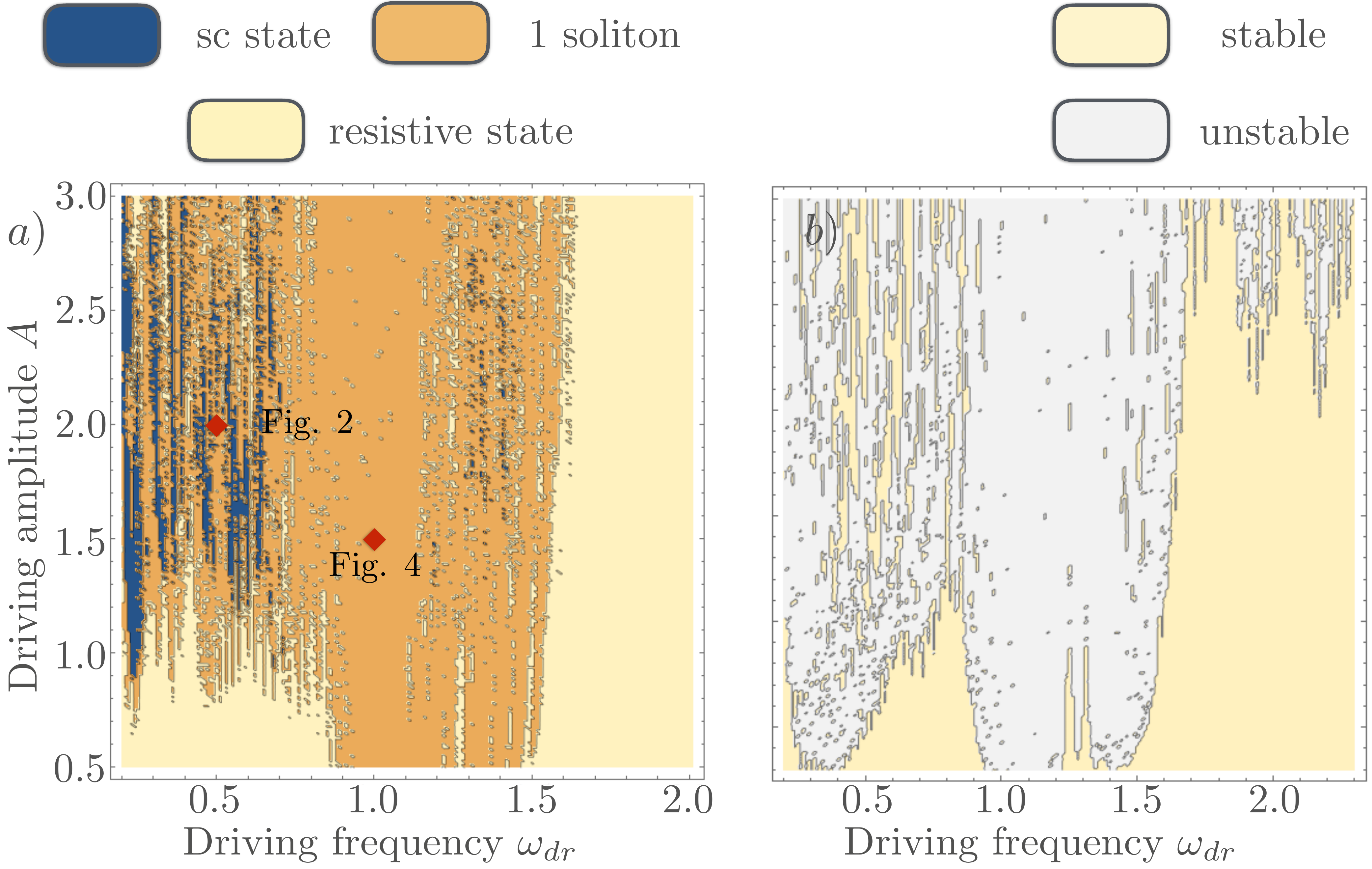}
\caption{
a) Final macroscopic quantum state after excitation by a pulse with inverse bandwidth $\sigma = 10$, amplitude $A$, and carrier frequency $\omega_{dr}$. The system is initialised in the resistive state. 
 b) Destabilisation of the COM mode according to Eqs.~(\ref{eq.eigenmodes_res1})-(\ref{eq.eigenmodes_res3}).}
\label{fig.parameter-space2}
\end{figure}

\section{Parameter dependence of the final states \& switching mechanism}
\label{sec.parameter}
We now investigate systematically the parameter space spanned by the driving frequency $\omega_{dr}$ and the amplitude $A$ in Eq.~(\ref{eq.pulse-form}).
This will allow us to explain the mechanism underlying the switching between different states. We first discuss the destabilization of the sc state, then of the resistive state, and finally of the soliton.

\subsection{Destabilisation of the sc state}
Fig.~\ref{fig.parameter-space}a) shows the final state of the system after driving the sc state by a pulse with amplitude $A$ and frequency $\omega_{dr}$. 
{We identify three resonances that destabilise the sc state at low driving strengths and define two distinct regions:}
For $\omega_{dr} \leq 1.5$, solitonic states are excited predominantly, with tiny islands of resistive and sc states in between. In contrast, in the region $\omega_{dr} \gtrsim 1.8$ the pulses excite solely resistive states (as exemplified in Fig.~\ref{fig.time-evolution1}, which falls into this region of parameter space). While the boundary between the latter region and the sc region appears regular, the former region is fairly irregular, with closely intertwined sc and solitonic solutions. 
Despite these irregularities, Fig.~\ref{fig.parameter-space} demonstrates that wide regions in parameter space exist in which solitonic or resistive states can be excited deterministically. 

Next we establish a simple model for the explanation of the results shown in Fig.~\ref{fig.parameter-space}a). We derive the analytic solution of the linearised equation of motion for the phase in appendix~\ref{appendix-EOM}, and find that it exhibits pronounced peaks at the eigenmodes $k_n = n \pi / L$ of the undriven system. Therefore, to gain a better understanding {of} the structure in Fig.~\ref{fig.parameter-space}a), we make the ansatz 
\begin{align}
\phi (\xi, \tau) &= \phi_{\text{sc}} +\sum_n f_n (\tau) \cos \left( \frac{n \pi \xi}{L} \right), 
\end{align}
where we recall the sc state $\phi_{\text{sc}} = \arcsin (j_{\text{ext}})$, and expand the Josephson coupling to 
leading nonlinear order, $\sin (\phi_{\text{sc}} + \delta ) \simeq \sin (\phi_{\text{sc}}) + \cos (\phi_{\text{sc}}) \delta - \sin (\phi_{\text{sc}}) \delta^2 / 2$. 
The linear order $\cos (\phi_{\text{sc}}) \delta$ merely yields decoupled wave equations for the $k$-modes with the potential $[ (1 - j_{\text{ext}}^2)^{1/2} + k_n^2 ] f_n^2 / 2$. The next order, $\sin (\phi_{\text{sc}}) \delta^2 / 2$, has two effects: 
First, it couples the various eigenmodes' equations of motion{, such that the resonant excitation of a specific mode can influence the system at all frequencies.}
Second, it also changes the potential of the centre-of-mass (COM) mode $n = 0$ to a cubic one, $(1 - j_{\text{ext}}^2)^{1/2} f_0^2 / 2 - j_{\text{ext}} f_0^3 / 3$, 
which features a stable equilibrium point at $f^{\ast}_0 = 0$ as before, but further adds an unstable one at $f_0^{\ast \ast} = (1 - j_{\text{ext}}^2)^{1/2} / j_{\text{ext}}$. 
When the excitation exceeds this point{,} the dynamics would become unbounded to leading nonlinear order. This light-induced destabilisation of fluctuations around the steady state signifies the possible switching to a different macroscopic quantum state.
The COM equation of motion reads
\begin{align}
&f_0'' (\tau) + \nu_c f_0' (\tau) + \sqrt{1 - j_{\text{ext}}^2} f_0 (\tau) - \frac{j_{\text{ext}}}{2} f_0^2 (\tau) 
= f_{dr} (\tau), \label{eq.com-mode}
\end{align}
where $f_{dr} (\tau)$ describes the driving of the COM mode. As discussed above, {the driving $f_{dr}$} consists of two contributions - the direct excitation by the light pulse, as well as {the indirect excitation through }the nonlinear coupling to other modes [see Eqs.~(\ref{eq.eigenmodes1}) to (\ref{eq.eigenmodes3}) {for details}]. We model the external driving as identical to the external pulse, $A \sin (\omega_{dr} \tau) \exp [- \tau^2 / (2 \sigma^2)] $. This does not capture details of the pulse propagation in the system, {and deviations must be expected - in particular at large driving amplitudes, where the full system is expected to saturate. 
Nevertheless it} is sufficient to understand the main physical properties of the full model, as shown below. 

As shown in Fig.~\ref{fig.parameter-space}b), the simple model Eq.~(\ref{eq.com-mode}) is able to reproduce the different parameter regimes shown in panel a) very well. While it cannot reproduce the irregular speckle-like patterns below $\omega_{dr} < 0.5$ of the full model, it does correctly reproduce the three resonances. The deviation at low frequencies originates from the simplified model allowing excitations at low frequencies while the full model predicts almost complete reflection, with excitations only being permitted above the nonlinear supratransmission threshold \cite{Geniet02}.
{The simplified model} further allows {associating} the resonances with the {resonant} excitation of the k-modes:
The lowest-energy resonance stems predominantly from the direct excitation of the $n = 0$ (COM) mode. 
The other two resonances originate from indirect excitation via the quadratic coupling to the $n = 1$ mode 
and to the $n=2$ mode
, respectively. 

The good agreement between Figs. \ref{fig.parameter-space}a) and b) demonstrates that the switching between macroscopic states may be understood as the light-induced destabilisation of the COM mode. 
Our simple model gives rise to correct predictions not only for the excitation of the uniform resistive state, but also for the highly localised moving soliton state where intuition could suggest a close connection to high-$k$ modes. 
However, our explanation is insufficient to predict the final state following the destabilisation.  

\subsection{Destabilisation of the resistive state}

The parameter space for the destabilisation of the resistive state is shown in Fig.~\ref{fig.parameter-space2}a).
It demonstrates that the resistive state can be destabilised by low-frequency irradiation with driving frequency $\omega_{dr} \lesssim 1.5$, while it remains stable against high-frequency pulses. We again compare these simulations with {results from} a simplified model, which we derive in Eqs.~(\ref{eq.eigenmodes_res1})-(\ref{eq.eigenmodes_res3}), shown in panel b). Just like in the case of the sc state, it overestimates the instability at very low frequencies, $\omega_{dr} \lesssim 0.5$, and further predicts instabilities above $\omega_{dr} \gtrsim 1.5$. The latter can be explained by an overestimation {of} the maximal Josephson current at large driving: Since we introduce the driving directly into the equations of motion, {the excitation increases linearly with the amplitude $A$.}
Due to the nonlinear Josephson coupling, this is not the case in the full dynamics, and the simplified model thus overestimates the instability. 
Yet, it correctly predicts the transition frequency $\omega_{dr} \sim 1.5$, as well as {the} comparatively large amplitude needed to destabilize the resistive state for $\omega_{dr} \lesssim 0.5$. {Like for the sc state, }
it is the destabilisation of the centre-of-mass mode that is responsible for the switching between macroscopic states. 

{Although }the destabilisation mechanism is the same, the {resulting final state structures in the} parameter space are vastly different compared to the sc state in Fig.~\ref{fig.parameter-space}. 
{This is due to the different dispersion relations, which is linear, $\omega = |k|$, in the resistive state, and in contrast features a band gap,  $\omega ^2 = \sqrt{1 - j_{\text{ext}}^2} + k^2$, in the sc state.}
Hence, the all-important centre-of-mass mode shifts to zero frequency, rendering the resistive state susceptible to low-frequency driving, and stable against high-frequency perturbations.

Furthermore, we remark that both the results from the simulations and those from the simplified model are highly sensitive on the driving frequency $\omega_{dr}$. A small change in the driving frequency can result in a different final state (or change the dynamics from stable to unstable in the simplified model), whereas small changes of the driving amplitude seldom change the dynamics. As we will discuss next in the context of the destabilization of the soliton, whose parameter space shows similar features, the use of single- or few-cycle pulses creates more regular structures, since their larger bandwidth effectively averages over a frequency interval, resulting in larger intervals with a unique final state. 


\subsection{Destabilisation of the soliton}

\begin{figure}
\centering
\includegraphics[ width=0.3\textwidth]{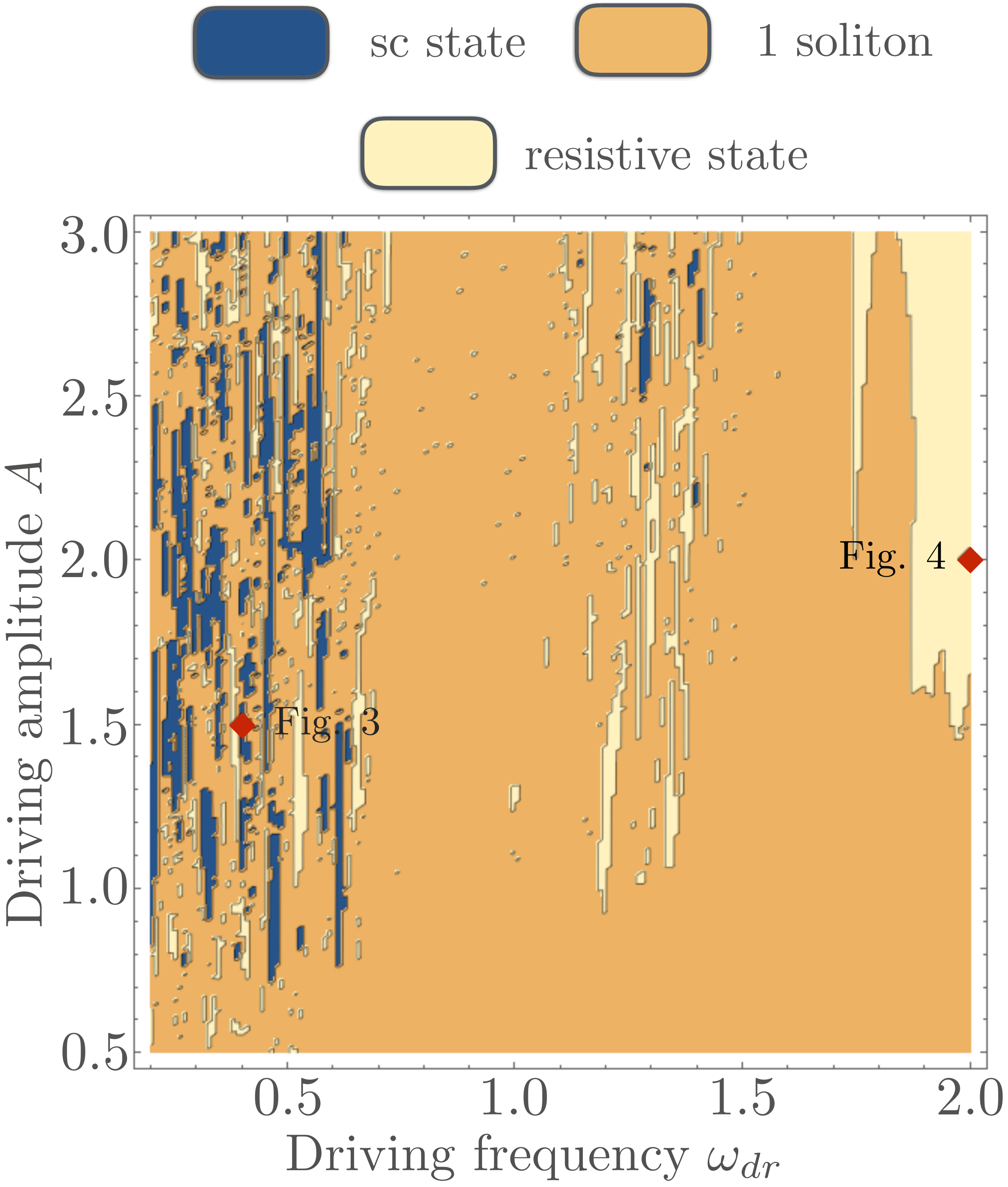}
\caption{
Final macroscopic quantum state after excitation by a pulse with inverse bandwidth $\sigma = 10$, amplitude $A$, and carrier frequency $\omega_{dr}$. The system is initialised in the soliton state. 
 }
\label{fig.parameter-space3}
\end{figure}


Fig.~\ref{fig.parameter-space3} presents the final state after the interaction with a pulse with constant pulse duration $\sigma = 10$ ($i.e.$ as before). It demonstrates that it is in fact possible to destroy the soliton, and reset the system {to} the sc state (blue regions) or the resistive state (bright yellow regions). The soliton is mostly vulnerable against low-frequency driving with $\omega_{dr} \lesssim 0.6$, but also against very high frequency excitation with $\omega_{dr} \gtrsim 1.8$. Yet, the structure appears {irregular, and shows the same vertical structure along the vertical axis we observed in Fig.~\ref{fig.parameter-space2}, where small changes of $\omega_{dr}$ can change the final state.}  
With the exception of {driving at} very high frequencies - one cannot identify large regions{, where the soliton can be destroyed with confidence.}

\begin{figure*}
\centering
\includegraphics[width=0.8\textwidth]{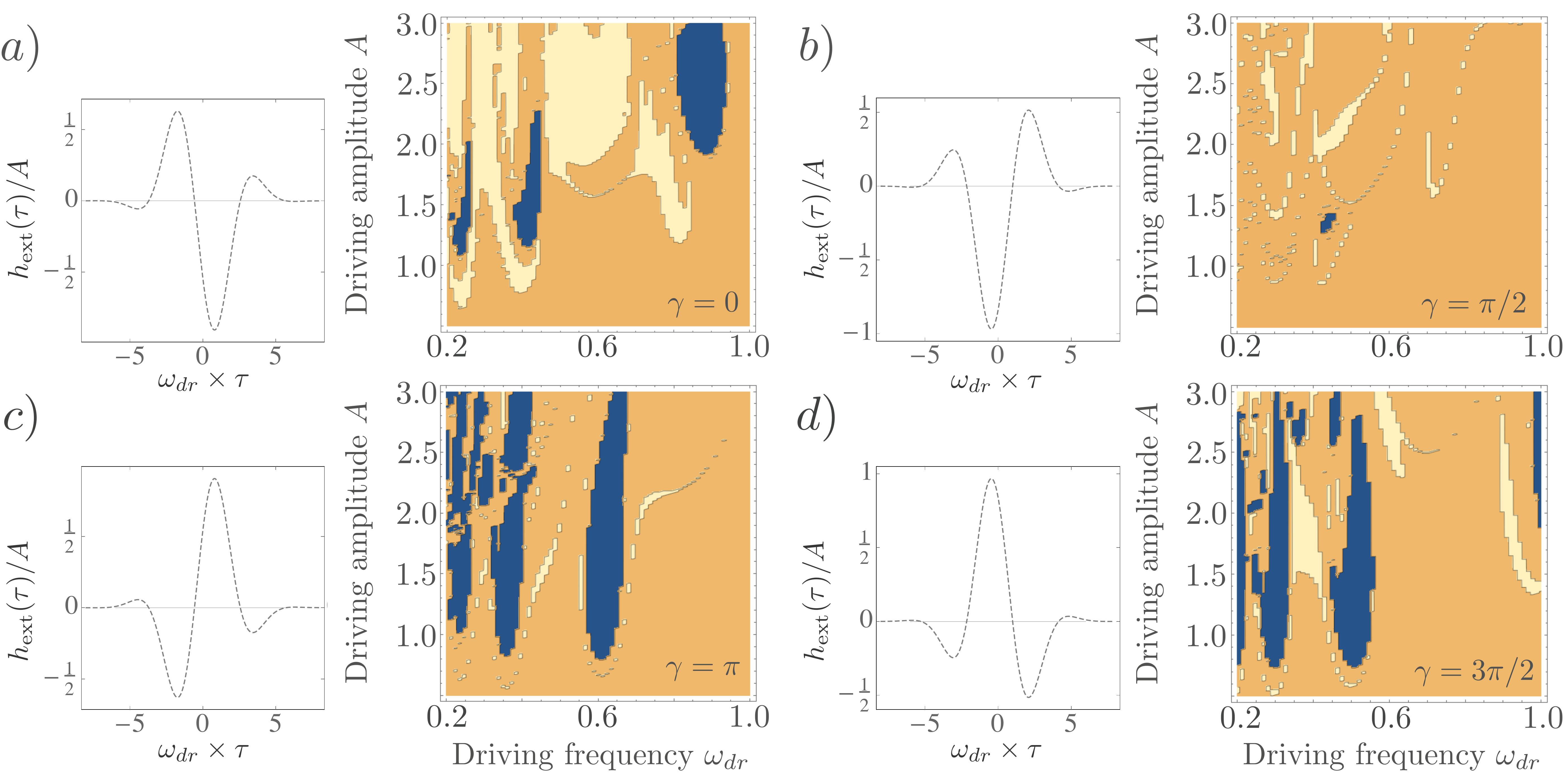}
\caption{
Final macroscopic quantum state after excitation by
{ a single cycle pulse with bandwidth $\sigma = 2 / \omega_{dr}$ and CEP a) $\gamma = 0$, b) $\pi / 2$, c) $\pi$, and d) $3 \pi /2$. The system is initialised in the soliton state (same as Fig.~\ref{fig.parameter-space3}). The panels on the left indicate the pulse form for the given CEP, which is identical for any frequency.}
 }
\label{fig.parameter-space4}
\end{figure*}

As the soliton is a strongly localised wave, its interaction with short pulses differs from the previous two cases, in that the CEP $\gamma$ in Eq.~(\ref{eq.pulse-form}) may become important. Thus, if we instead simulate the interaction with single cycle pulses of duration $\sigma = 2 / \omega_{dr}$, we obtain the results shown in Fig.~\ref{fig.parameter-space4}. 
In these plots, we also pick $\tau_0 = 10 / \omega_{dr}$ to assure that pulses with different driving frequencies have the same shape, which is shown on the right side of the panels. {The oscillation frequency of the soliton $\omega_{\infty}$ is considerably larger than the driving frequency $\omega_{dr}$, such that the value of $\tau_0$ is not central to the results.}
We find that the larger bandwidth at lower frequencies averages out the irregular structure of Fig.~\ref{fig.parameter-space3}. This creates regions in which the switching can be accomplished with confidence. {In contrast to the earlier results in Fig.~\ref{fig.parameter-space3}, the final state after excitation by these short pulses strongly depends on the phase. In our simulation, the phase $\gamma = 0$ in panel a) favours the excitation of the resistive state. While at intermediate values in panel b) it can only seldom destroy the soliton, at larger values in panel c) it favours the excitation of the sc state.

This behaviour can be understood qualitatively through the analysis of the pulse form, which is shown for each case on the left of the panels. Note that for each set of parameters $\{ \omega_{dr}, A \}$, this pulse is stretched or compressed in both time and amplitude, but it always retains this shape.
At $\gamma = 0$, negative values {of the field amplitude} dominate the pulse form. A negative magnetic field at the left boundary implies that $\partial h / \partial \xi > 0$, as long as no other magnetic fields are present. In Eq.~(\ref{eq.EOM1}), this lowers the {value of the} left-hand side of the equation. Thus, the quasiparticle current term $\nu_c \partial \phi / \partial \tau$ has to rise, such that the sum of the terms equals the external current. 
Conversely, at $\gamma = \pi$, positive values dominate in the pulse, thus reducing the instantaneous voltage, and thereby favouring the excitation of the sc state.

\section{Discussion \& Conclusions}
\label{sec.discussion}

In conclusion, we have proposed to manipulate the macroscopic quantum state of a current-carrying layered superconductor using THz pulses. By focusing on the interaction with strong, few-cycle pulses, we investigated their use as ultrafast switches that can reset the system from the zero-voltage sc state to a finite-voltage state (either solitonic or resistive) and vice versa. We showed that this manipulation is enabled by the strong nonlinearity of the light-matter interaction in the system by means of a simple toy model. Here, the nonlinearity results in a driving term for the centre-of-mass mode of the plasma oscillations, whose destabilisation indicates the switching between macroscopic states. We have pointed out possible applications of these findings.

Driving the system below the plasma resonance does not affect the sc state below the supratransmission threshold \cite{Geniet02}, but it can destabilise the resistive or solitonic state. For instance, both pulses shown in Fig.~\ref{fig.time-evolution1}c) and \ref{fig.time-evolution2}b) do not destabilise the sc state, yet they can destabilise the resistive state, and thereby prohibit phase fluctuations from destroying the coherence between junctions. Our work thus points towards an unusual, yet feasible approach to the ongoing effort to laser cool superconducting fluctuations \cite{Sam15,Hoeppner15, Okamoto16}. 
{ Similarly, it will be interesting to explore other parameter regimes supporting different macroscopic states, and investigate, for instance, whether driving can destroy or stabilise vortex lattices in the presence of external magnetic fields.
}
This will be pursued in future work.

\begin{acknowledgments}
D.~J. thanks the Graduate School of Excellence Material Science in Mainz for hospitality during part of this work. 
The research leading to these results has received funding from the European Research Council under the European Union's Seventh Framework Programme (FP7/2007-2013) Grant Agreement No. 319286 Q-MAC. 
M.~K. thanks the National Research Foundation and the Ministry of Education of Singapore for support.
\end{acknowledgments}

\begin{appendix}
{
\section{Physical units}
\label{appendix.units}

The $c$-axis electrodynamics of layered superconductors is determined by the dielectric constant $\epsilon$, and three characteristic length scales: the penetration depths $\lambda_c$ along its $c$-axis, and $\lambda_{ab}$ along the $ab$-planes, as well as the interlayer spacing $s$. 
From these, we can construct the Josephson plasma frequency
\begin{align}
\omega_p &= \frac{c}{\sqrt{\epsilon} \lambda_c},
\end{align}
the number of magnetically coupled junctions, 
\begin{align}
\ell &= \frac{\lambda_{ab}}{s},
\end{align}
and the anisotropy parameter $\gamma = \lambda_{ab} / \lambda_c$. All currents in this paper are normalized to the critical Josephson current, which is given by $j_J = c \Phi_0 / (8 \pi^2 s \lambda_c)$ \cite{Koshelev01}. 
Time is measured in units of the plasma frequency, $i.e.$ $\tau = \omega_p \times t$, and the spatial coordinate in units of the c-axis penetration length $\xi = x / \lambda_c$. 
Furthermore, the quasiparticle conductivities along the $c$-axis $\sigma_c$ and the $ab$-plane $\sigma_{ab}$ are converted into dimensionless damping rates by the relations
\begin{align}
\nu_c &= \frac{4 \pi \sigma_c}{\epsilon \omega_p}, \\
\nu_{ab} &= \frac{4 \pi \sigma_{ab}}{\epsilon \omega_p \gamma^2}.
\end{align}
The dimensionless magnetic field $h_n$ is measured in units of $B_{0} = \Phi_0 / (2 \pi \lambda_c s)$.
The electric field is given by \cite{Koshelev01}
\begin{align}
E_z &= \frac{\Phi_0}{ 2 \pi c s} \frac{\partial \phi_n}{\partial t} \\
&= \frac{B_0}{\sqrt{\epsilon}} \frac{\partial \phi_n}{\partial \tau}.
\end{align}
Gaussian units are employed.


}

\section{Centre-of-mass mode dynamics}
\label{appendix-EOM}

Here we derive the simplified equations of motion of eigenmode fluctuations around a given steady state of the system $\phi_0$. To this end, we write the full wavefunction as a sum,
\begin{align}
\phi (\xi, \tau) &= \phi_0 + \phi_{\epsilon} (\xi, \tau),
\end{align}
where $\phi_0$ denotes a dynamical steady state solution of the full mode, {which could be either the sc state $\phi_{\text{sc}}$, the resistive state~(\ref{eq.resistive}), or the soliton~(\ref{eq.soliton}),} 
and $\vert \phi_{\epsilon} \vert \ll \vert \phi_0 \vert$. 
Inserting this ansatz into the sine-Gordon equation, 
we obtain to second order
\begin{align}
\frac{\partial^2 \phi_{\epsilon}}{\partial \tau^2} + \nu_c \frac{\partial \phi_{\epsilon}}{\partial \tau} + \cos (\phi_0) \phi_{\epsilon}  - \frac{\partial^2 \phi_{\epsilon}}{\partial \xi^2} &= \frac{1}{2} \sin (\phi_0) \phi_{\epsilon}^2. \label{eq.fluctuations}
\end{align}
The left-hand side of Eq.~(\ref{eq.fluctuations}) describes the propagation of linear waves, with the system in the stated state $\phi_0$. The right-hand side yields corrections to this wave behaviour at larger amplitudes.

\subsection{Linear waves}

\subsubsection{sc state}
In the sc state, we have $\cos (\phi_{\text{sc}}) = (1 - j_{\text{ext}}^2)^{1/2}$ and $\sin (\phi_{\text{sc}}) = j_{\text{ext}}$. In {this} case, the left-hand side of Eq.~(\ref{eq.fluctuations}) represents a linear wave equation which we can solve straightforwardly {by Fourier transform}. We seek a solution of the linearised sine-Gordon equation,
\begin{align}
\frac{\partial^2 \phi_{\epsilon}}{\partial \tau^2} + \nu_c \frac{\partial \phi_{\epsilon}}{\partial \tau} + \sqrt{1 - j_{\text{ext}}^2} \phi_{\epsilon} - \frac{\partial^2 \phi_{\epsilon}}{\partial \xi^2} &= 0, \label{eq.linear-de}
\end{align}
subject to the boundary conditions
\begin{align}
\frac{\partial \phi_{\epsilon}}{\partial \xi} \bigg\vert_{\xi = 0} &= h_{ext} (\tau), \label{eq.left-boundary} \\
\frac{\partial \phi_{\epsilon}}{\partial \xi} \bigg\vert_{\xi = L} &= 0. \label{eq.right-boundary}
\end{align}
with $h_{ext}$ given by Eq.~(\ref{eq.pulse-form}). Writing the wavefunction as
\begin{align}
\phi_{\epsilon} (\xi, \tau) &= \int \!\! d\omega \; \Im e^{i \omega \tau} \phi_{\omega} (\xi), 
\end{align}
with 
\begin{align}
\phi_{\omega} (\xi) &= A \cos k_{\omega} \xi + B \sin k_{\omega} \xi,
\end{align}
we obtain from Eq.~(\ref{eq.linear-de}) the dispersion relation
\begin{align}
\omega^2 - i \nu_c \omega &= \sqrt{1 - j_{\text{ext}}^2} + k_{\omega}^2, \label{eq.sc-dispersion}
\end{align}
and from the boundary conditions~(\ref{eq.left-boundary}) and (\ref{eq.right-boundary}),
\begin{align}
B &= \frac{h_{ext} (\omega)}{2 \pi k_{\omega}}, \\
A &= B \tan^{-1} k_{\omega} L.
\end{align}
Thus, we arrive at
\begin{align}
\phi_{\epsilon} (\xi, \tau) &= \int \!\!\! d\omega \; \Im e^{i \omega \tau} \frac{h_{ext} (\omega)}{2 \pi k_{\omega}} \left[ \tan^{-1} (k_{\omega} L) \cos (k_{\omega} \xi) + \sin (k_{\omega} \xi)  \right]. \label{eq.linear-solution}
\end{align}
Clearly, this solution is strongly peaked whenever $k_{\omega} = n \pi / L$, $i.e.$ when the driving excites a cavity resonance in the junction. 

\subsubsection{resistive state}
We approximate the resistive state~(\ref{eq.resistive}) by its dominant term $\phi_{\text{resistive}} \simeq \omega_0 \tau$ to obtain the Mathieu equation 
\begin{align}
\frac{\partial^2 \phi_{\epsilon}}{\partial \tau^2} + \nu_c \frac{\partial \phi_{\epsilon}}{\partial \tau} + \cos (\omega_o \tau) \phi_{\epsilon} - \frac{\partial^2 \phi_{\epsilon}}{\partial \xi^2} &= 0. \label{eq.Mathieu}
\end{align}
With the parameters employed in this manuscript, we have $\omega_0 = 2.5$ - hence, it is much larger than frequencies around and below the plasma resonance with $\omega \leq 1.0$, that we are most interested in. Therefore, there will be very little mixing between these widely disparate frequencies, and in a first approximation, we replace the highly oscillatory term $\cos (\omega_0 \tau)$ by its time-averaged value $= 0$. This creates a free space-like wave equation with linear dispersion,
\begin{align}
\omega^2 - i \nu_c \omega  &= k^2_{\omega}.
\end{align}
Whereas Eq.~(\ref{eq.sc-dispersion}) only supports (approximately) real wavevectors above the plasma edge, when $\omega > (1 - j^2_{\text{ext}})^{1/2}$, there is no forbidden spectral region in the resistive state, and low-frequency waves can penetrate the system.

\subsubsection{solitonic state}
In an infinitely long junction, a soliton breaks the time-transversal symmetry. This creates a Goldstone mode at zero frequency, while the remaining dispersion relation is not affected by the presence of the soliton \cite{Fogel77}. Hence, the existence of the zero-frequency mode can explain the susceptibility of the soliton state against low-frequency driving.

\subsection{Destabilisation of the sc state}

{The linear wave dispersion~(\ref{eq.sc-dispersion}) }
motivates us to reduce our description to only the first few eigenmodes with $n = 0,1, 2$ in the investigation of the nonlinear corrections to the solution~(\ref{eq.linear-solution}) {[This is also supported by the reflectivity spectrum in Fig.~\ref{fig.reflectivity}a)]}. We write 
\begin{align}
\phi_{\epsilon} (\xi, \tau) &= \sum_{n = 0}^{2} f_n (\tau) \cos \left( \frac{n \pi \xi}{L} \right),
\end{align} 
multiply Eq.~(\ref{eq.fluctuations}) by $\cos (n \pi x / L)$, and integrate over space. This approach results in the coupled equations of motion of the eigenmodes:
\begin{align}
f_0'' (\tau) + &\nu_c f_0' (\tau) + \sqrt{1 - j_{\text{ext}}^2} f_0 (\tau) - \frac{j_{\text{ext}}}{2} f_0^2 (\tau) \notag \\
= &\frac{j_{\text{ext}}}{4} \left[ f_1^2 (\tau) + f_2^2 (\tau) \right] + f_{0 dr} (\tau), \label{eq.eigenmodes1} \\
f_1'' (\tau) + &\nu_c f_1' (\tau) + \left[ \sqrt{1 - j_{\text{ext}}^2} + \left( \frac{\pi}{L} \right)^2 \right] f_1 (\tau) \notag \\
= & j_{\text{ext}} \left[ f_0 (\tau) f_1 (\tau) + \frac{1}{2} f_1 (\tau) f_2 (\tau)  \right] + f_{1 dr} (\tau), \\
f_2'' (\tau) + &\nu_c f_2' (\tau) + \left[ \sqrt{1 - j_{\text{ext}}^2} + \left( \frac{2 \pi}{L} \right)^2 \right] f_2 (\tau) \notag \\
= & j_{\text{ext}} \left[ f_0 (\tau) f_2 (\tau) + \frac{1}{4} f_1^2 (\tau)  \right] + f_{2 dr} (\tau). \label{eq.eigenmodes3}
\end{align}
We have neglected the coupling to higher-$n$ modes, and added phenomenological driving terms which we simply write as
\begin{align}
2 f_{0 dr} (\tau) = f_{1 dr} (\tau) = f_{2 dr} (\tau) &= A \sin (\omega_{dr} \tau) e^{- (\tau - \tau_0)^2 / (2 \sigma^2)},
\end{align}
$i.e.$ we assume that they have the same shape as the external driving~(\ref{eq.pulse-form}). The additional factor 2 ahead of $f_{0 dr}$ stems from the spatial integration, since $\int d\xi \; 1 = L$ and $\int dx \cos^2 (k_n x) = L / 2$, thus giving greater weight to the finite-$k$ modes. We stress that this approach neglects the details of the wavepacket propagation inside the Josephson junctions, as these details are not essential for the understanding of the switching process. 

\subsection{Destabilisation of the resistive state}

{As in the Eq.~(\ref{eq.Mathieu}), we approximate the resistive state~(\ref{eq.resistive}) by its dominant term $\phi_{\text{resistive}} \simeq \omega_0 \tau$. }Inserting {the approximation} into Eq.~(\ref{eq.fluctuations}), the same approach as above yields
\begin{align}
f_0'' (\tau) + &\nu_c f_0' (\tau) + \cos( \phi_{\text{res}} (\tau) ) f_0 (\tau) - \frac{\sin (\phi_{\text{res}} (\tau)) }{2} f_0^2 (\tau) \notag \\
= &\frac{\sin ( \phi_{\text{res}} (\tau) )}{4} \left[ f_1^2 (\tau) + f_2^2 (\tau) \right] + f_{0 dr} (\tau), \label{eq.eigenmodes_res1} \\
f_1'' (\tau) + &\nu_c f_1' (\tau) + \left[ \cos (\phi_{\text{res}} (\tau)) + \left( \frac{\pi}{L} \right)^2 \right] f_1 (\tau) \notag \\
= & \sin ( \phi_{\text{res}} (\tau)) \left[ f_0 (\tau) f_1 (\tau) + \frac{1}{2} f_1 (\tau) f_2 (\tau)  \right] + f_{1 dr} (\tau), \\
f_2'' (\tau) + &\nu_c f_2' (\tau) + \left[ \cos ( \phi_{\text{res}} (\tau) ) + \left( \frac{2 \pi}{L} \right)^2 \right] f_2 (\tau) \notag \\
= & \sin ( \phi_{\text{res}} (\tau) ) \left[ f_0 (\tau) f_2 (\tau) + \frac{1}{4} f_1^2 (\tau)  \right] + f_{2 dr} (\tau). \label{eq.eigenmodes_res3}
\end{align}
{
We write the driving fields as
\begin{align}
4 f_{0 dr} (\tau) = 2 f_{1 dr} (\tau) = 2 f_{2 dr} (\tau) &= A \sin (\omega_{dr} \tau) e^{- (\tau - \tau_0)^2 / (2 \sigma^2)},
\end{align}
where the factor $2$ is inserted to roughly match the simulations of the full model.
}

\section{Fluctuations}
\label{sec.fluctuations}

In the absence of external currents, the only ground state is the sc state. We therefore expand the random driving in eigenmodes of the sc states,
\begin{align}
\eta (\xi, \tau) &= \sum_n \epsilon_n (\tau) \cos (k_n \xi) \label{eq.noise-expansion}
\end{align}
Following the fluctuation-dissipation theorem, we assume the power spectrum of the form
\begin{align}
\big\langle \epsilon_n (\tau) \epsilon_{n'} (\tau') \big\rangle &= \frac{\alpha k_B T}{\omega_n} \; \delta_{n n'} \delta (\tau - \tau'),
\end{align}
with the frequencies $\omega_n = ( (1 - j_{\text{ext}}^2)^{1/2} + k_n^2)^{1/2}$, and a proportionality factor $\alpha$. This corresponds to so-called ``pink noise", where high-frequency fluctuations are suppressed \cite{Weissman}\footnote{We also checked numerically the influence of white noise,  $i.e.$ $\big\langle \epsilon_n (\tau) \epsilon_{n'} (\tau') \big\rangle = \alpha k_B T \; \delta_{n n'} \delta (\tau - \tau')$, with a sharp cut-off at $\omega = 10$. It didn't affect the results of the time evolutions in this publication. {Generally, we note that the noise can slightly move the boundaries in Figs.~\ref{fig.parameter-space} - \ref{fig.parameter-space3}, $i.e.$ it can influence the switching near criticality, but it does not affect the larger structure of the plots.}
}.

In our simulations, we create a single realisation of random functions $\{ \epsilon_n (t) \}${, which we interpolate from random values drawn from a Gaussian distribution with zero mean and variance $\alpha k_B T / \omega_n$ with a temporal step size $\Delta\tau$ set to one}, and $\alpha k_B T = 0.005$. {The summation in Eq.~(\ref{eq.noise-expansion}) is terminated at $n = 10$, corresponding to a frequency cut-off $\omega_{10} \sim 9.5$ for our parameters. }
{This function} is then fed into the equation of motion~(\ref{eq.EOM1}) where it acts as a random scattering potential. 
Its effect is shown in Fig.~\ref{fig.reflectivity}, where we calculate the reflectance $| r (\omega) |^2$ from the propagation of a plasma wave.

Panel a) shows the signal from a short junction with length $L = 3.3$ without noise. The left plot shows the time evolution of the external field $E_{ext}$ (grey) and the reflected field $E_r$ (red). This is used to obtain the reflectance $| r (\omega) |^2$, with $r = E_r / (E_r - 2 E_{ext})$. The reflectance features clear resonances at the eigenmodes of the junction with wavevectors $k_n = n \pi / L$ with n = 0,1, 2. 

Panel b) shows the same simulations with finite fluctuations. The reflectance still has the same features, the noise merely adds random fluctuations on top of the signal.

Finally, in panel c) we present the simulations without noise in a long junction with $L = 100$. The eigenmodes now overlap entirely, and the reflectance perfectly coincides with the theoretical expectation, which we obtain from the frequency-dependent dielectric constant \cite{Koshelev07}
\begin{align}
\epsilon (\omega) &= \epsilon_0 \left( \sqrt{ 1 - j_{\text{ext}}^2 } - \frac{1}{\omega^2} + i \frac{\nu_c}{\omega} \right). \label{eq.epsilon}
\end{align}

\begin{figure*}
\centering
\includegraphics[width=0.8\textwidth]{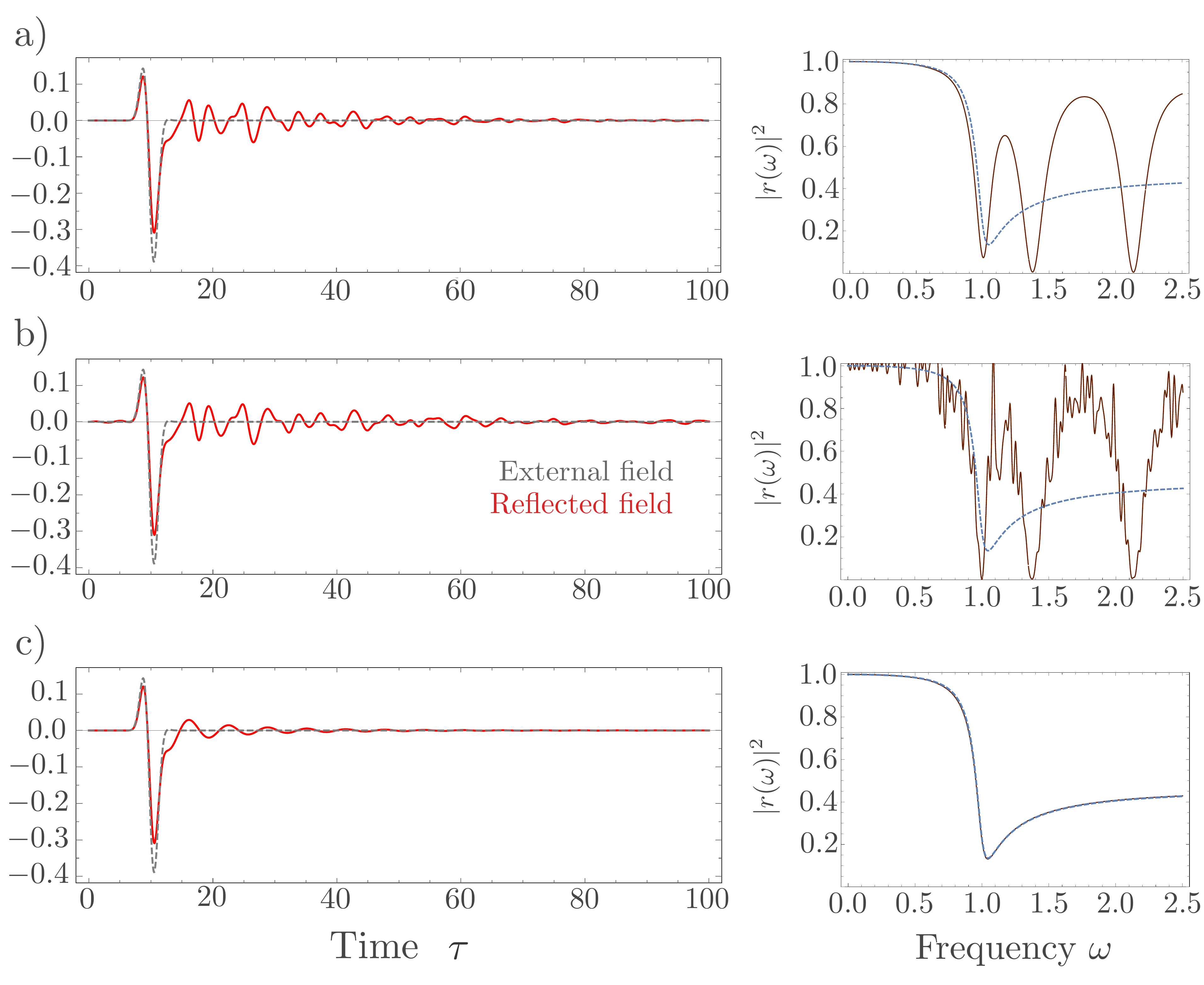}
\caption{\textbf{Optical signals and noise:} a) Time evolution (left) and reflectance $| r (\omega) |^2$ (right) of a weak excitation in a short junction without noise. The blue, dashed line in the reflectance plot shows the theoretical expectation for the bulk system according to Eq.~(\ref{eq.epsilon}).
b) The same including the noise level used in the paper.
c) Time evolution and reflectance in a long junction. The simulations and theoretical expectation coincide, and cannot be distinguished.
}
\label{fig.reflectivity}
\end{figure*}

\end{appendix}

\end{document}